# On the Scientific Value of Large-scale Testbeds for Wireless Multi-hop Networks

Mesut Güneş
mesut.guenes@ovgu.de
Communication and Networked Systems (ComSys)
Institute for Intelligent Cooperating Systems
Faculty of Computer Science
University of Magdeburg

## Abstract

Large-scale wireless testbeds have been setup in the last years with the goal to study wireless multi-hop networks in more realistic environments. Since the setup and operation of such a testbed is expensive in terms of money, time, and labor, the crucial question rises whether this effort is justified with the scientific results the testbed generates.

In this paper, we give an answer to this question based on our experience with the DES-Testbed, a large-scale wireless sensor network and wireless mesh network testbed. The DES-Testbed has been operated for almost 5 years. Our analysis comprises more than 1000 experiments that have been run on the testbed in the years 2010 and 2011. We discuss the scientific value in respect to the effort of experimentation.

## 1 Introduction

A *wireless multi-hop network* (WMHN) is a pure wireless network. In a WMHN two nodes can communicate directly if they are in their mutual transmission area. Otherwise intermedia nodes have to forward the packets hop-by-hop until the destination node is reached. The most prominent examples of WMHN are *wireless sensor networks* (WSN), *wireless mesh networks* (WMN), and *mobile ad hoc networks* (MANET).

A WMHN represents a classic wireless distributed network without a central control, sometimes with node mobility and churn of nodes. As usual in the field of computer networks, the study and performance evaluation of WMHN is done either by mathematical analysis, simulation studies, or by building a research facility consisting of hardware and software. The latter case we denote as a *testbed*.

Since several years we witness a transition to experimentally-driven research in the domain of wireless communications and particularly wireless multi-hop networks. The main ratio behind this transition is that results obtained via analytical studies or simulations do not hold in reality [1]. The studied models suffer from simplifications, which render the results mostly useless. This is particularly true for the physical layer, since modeling the radio propagation is very complex, especially in non-line of sight (NLOS) scenarios, when the influence of the environment has to be taken into account [2]. Many application scenarios of wireless multi-hop networks are either located in buildings or urban areas and thus most of the wireless links are assumed to be non-line of sight. Therefore, a lot of wireless testbeds have been deployed to study these networks in a more realistic environment in recent years [3].

While the results of testbed-based studies promise a higher degree of accuracy, the required effort in terms of money, time, and labour is significantly higher compared for example to experiments with simulation tools [4]. First of all, real hardware is required and on top of the hardware customised (open?[1]) system level software is needed, which can be used for scientific work. Furthermore, after the initial deployment, the testbed has to be actively maintained on a day-to-day basis.

In a large-scale testbed with hundreds of nodes, this effort involves usually several people and the testbed may be used in multiple projects (and probably funded by them) [5]. Thus, it is not surprising that one main question raised is the following:

> *Is the effort to develop and operate a large-scale wireless multi-hop network testbed justified with the results and scientific impact the testbed generates?*

In this paper we seek an answer to this question. Our answer is based on our experience with the DES-Testbed, a wireless sensor network and wireless mesh network testbed with more than 135 nodes at the Freie Universität Berlin, Germany [6]. The testbed has been operated for more than 5 years, thus, it is time to draw conclusions and critically question the scientific results achieved over the years. For this, we analyze all experiments run on the DES-Testbed during two years (2010 and 2011) in respect to the effort of the experiment, e.g., the experiment runtime, the number of nodes, and the scientific output. Based on these statistics, the raised question can clearly be answered in a positive way: *Yes, it was worth it.* However, the analysis reveals that the value and success of a large-scale testbed infrastructure is strongly related to its *management system* [7, 8]. Thus, it is a crucial requirement for the management system to provide tools that enforce the *scientific method* on the experimenters and provide appropriate tools for all steps of the *experiment life cycle* for networking experiments [9].

The remainder of this paper is structured as follows. In Section 2 we discuss the scientific method, its adaptation to the (wireless) networking domain, and we develop a scientific workflow for large-scale experimentation. In Section 3 we describe the DES-Testbed. In Section 4, we present statistics and quantitative data of our experience with the testbed. In Section 5 we conclude our findings.

## 2 The Scientific Method

In this section we review the scientific method for empirical research in the domain of networking and distributed systems and particularly consider wireless multi-hop networks. For this, we introduce a general *experiment life cycle* model and a *scientific workflow* for large-scale experimentation in wireless testbeds.

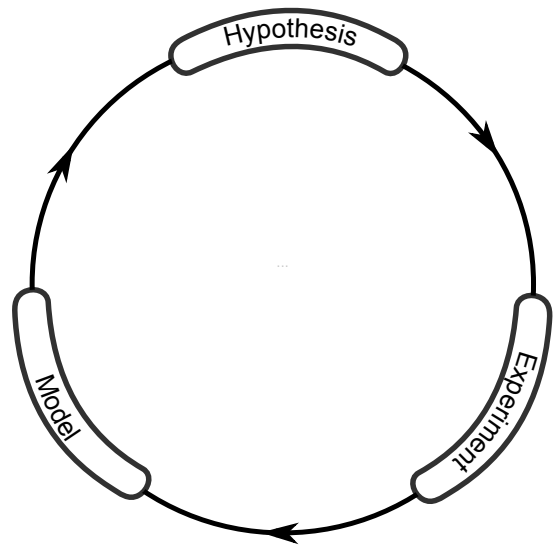

**Figure 1:** *The scientific method consisting of three activities, which are connected. Thus the research work has to be in a loop and repeated.*

The *scientific method* [9] is the fundamental approach in science to attain and expand knowledge. The basic idea is depicted in Figure 1 and consists of three activities. First, we formulate a *hypothesis* that is a suggested explanation of a certain behavior of the study subject. Second, we perform controlled *experiments* to collect data and gain insight into the studied subject. Third, we derive a *model* that is supported by the collected data. The results and the predictions of the model is then compared with the assumed hypothesis. The scientific method thus applies explanation-suggestions (hypothesis), controlled experiments, and observations to reduce uncertainty and to deduce and create new knowledge.

---

[1]As a scientist we wish to use mainly open source software, since it allows to understand the implementation of a technique, algorithm, or approach, which may influence the performed experiments and the obtained results.

As a consequence an experimental-study $S$ consists out of a sequence of *experiments* $E_1, E_2, \ldots, E_k$. Each experiment $E_i$ is repeated a specific number of times. Thus for experiment $E_i$ we perform $r_i$ *replications* and obtain the set of results $O_i = \{o_1, o_2, \ldots, o_{r_i}\}$. The scientific field of *design of experiments* (DoE) recommends to determine the number of replications $r_i$ based on the required accuracy of the results [10, 11].

In the remainder of this section we develop a scientific method to obtain $E_i$ for large-scale experiment series in a wireless multi-hop testbed.

## 2.1 Life cycle of an experiment

Figure 2 depicts the life cycle of an experiment in an abstract form. It does not contain any (wireless) networking specific elements. The experiment life cycle consists of three steps: (i) design, (ii) execution, and (iii) analysis.

### 2.1.1 Experiment design

In the first step of the experiment life cycle the experiment is developed. First the experiment idea and subsequently details of the experiment have to be defined precisely. The design of an experiment is a kind of art and the experimenter must have deep knowledge and expertise. For a networking experiment at least the following *parameters* have to be defined:

- Number of nodes

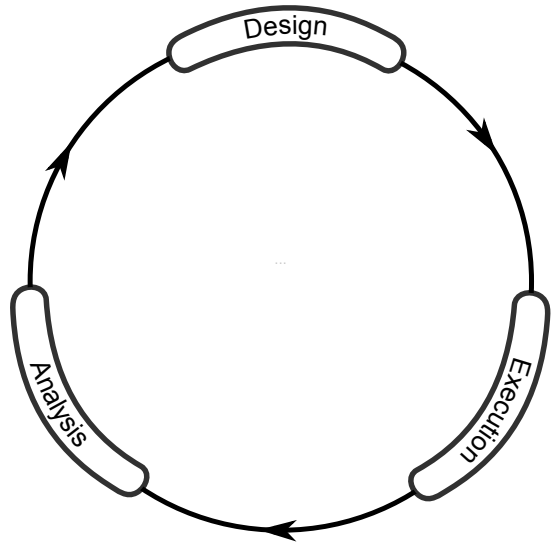

**Figure 2:** *Experiment life cycle consisting of three steps. The three steps match roughly the steps of the scientific method.*

- Topology of the network
- Nodes to be used, i.e., either static or dynamic selection of the used nodes.
- Roles of the nodes, i.e., client, server, or servent
- Traffic pattern
- Performance metrics, i.e., what is going to be measured and in which metric.

The set of the experiment parameters create the *parameter space* which can be huge. To improve the efficiency of the experimentation process, approaches from the field of *design of experiment* (DoE) should be used [10, 12]. By this the number of experiments and the number of replications can be reduced.

### 2.1.2 Experiment execution

In the second step the experiment will be executed. The most critical issue is that the experiment is started and finished in a *defined state*. This requires also that the measurements are started and finished in defined states. This is particularly important, since the data for the per-formance metrics are obtained by the measurements. Therefore this step should be performed by sub-steps. i) prepare the experiment system, ii) execute the experiment, and iii) cleanup the experiment system.

Another important and sensitive issue in this step is the *repeatability* of the experiment. There are two different forms of repeatability. Firstly, the repeatability of a whole experiment $E_i$ and secondly the repeatability of successive replications of an experiment. The assumption we want to comply with is the independency of the experiment runs for both cases, since most statistical methods used in the next step are based on this assumption.

### 2.1.3 Experiment analysis

In the last step of the experiment life cycle the observations are analyzed to obtain statistically sound data, which help to understand the occured processes during the experiment. For this *performance metrics* are developed for each experiment. To grasp the performance metrics quantitatively, point and interval estimators should be calculated in the analysis step.

We recommend to use boxplots to depict obtained results, which help to map a series of observation into a single plot [13].

## 2.2 Scientific workflow

Figure 3 depicts the scientific workflow adapted for the DES-Testbed, which is roughly oriented on [14] and consists of three steps: i) design, ii) execution, and iii) analysis. In this scheme the life cycle of an experiment is embedded into a workflow to support large-scale experimentation. In comparison to Figure 2, there is a new item in the center of the figure, *data*, which represents a general kind of a data storage, e.g., realized by a data base. The *data* is in fact manifold and represents different kinds of information that is created in the experiment life cycle and needs to be stored. The different steps of the experiment life cycle interact with *data* in different ways.

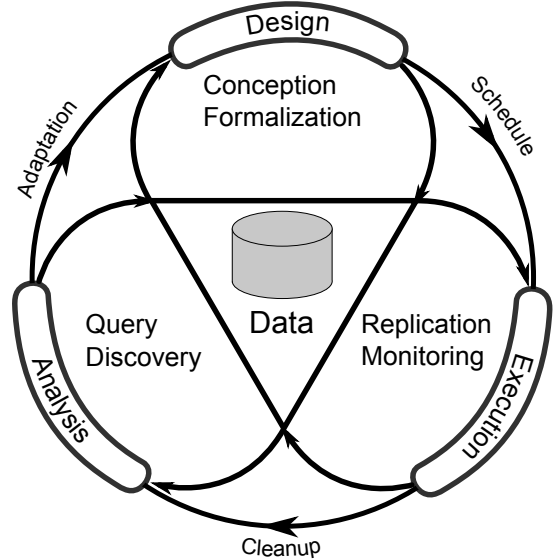

**Figure 3:** *Scientific workflow for empirical studies based on large scale experimentation. The workflow consists of three steps i) design, ii) execution, and iii) analysis. In the center of the workflow is a data storage, which stores all kinds of data related with the workflow, i.e., the experiment description, measurements, monitoring data, performance metrics, etc.*

### 2.2.1 Design

In the design step, the experiment idea has to be conceptualized and planned. Subsequently the concept of the experiment has to be formalized and transformed into a formal language that can be processed by computers. Domain specific languages (DSL) are appropriate for this task, since they provide the required expression tools. Advantages of this approach are:

- The formal representation of an experiment can be stored in the *data* storage for documentation and automatic experiment performance.

- Existing experiment descriptions can be reused to develop new experiments or to refine them.

- Formal experiment descriptions can be exchanged between scientist to redo and replicate experiments.

### 2.2.2 Execution

The execution step comprises two elements that are performed in parallel. Replication represents the repetition of experiments for a specific number of times. In an experiment we perform a measurement and obtain the experiment-data. Replications are an appropriate way to gain *experiment-data* that was generated independently. The experiment system is observed to gain *monitoring-data* besides the measurements during the experiment runtime. The experiment-data as well as the monitoring-data are stored in the data storage.

### 2.2.3 Experiment analysis

In this step the data is analyzed to get insights into the research subject. For this, the data has to be filtered, transformed, and the structural relationship has to be discovered. This step results in *performance metrics* that allow an interpretation of the experiment.

## 2.3 Requirements for testbeds

From the previous discussion of the experiment life cycle and the scientific workflow we distilled the following requirements for a specific implementation in a testbed:

- *Formalization:* The formalization of the experiments requires a formal experiment description language that is also capable to define start and end states for the experiment system.

- *Scheduling:* The support of large scale experiment series requires a scheduling system that can run experiments on planned schedules in an automated manner.

- *Replications:* The individual and independent repetition of experiments has to be performed between defined start and end states.

- *Monitoring:* A monitoring system should record the state of the experiment system. The monitoring system has to be independent of the measurements during the experiments to provide validation data.

- *Cleanup:* The cleanup is responsible to ensure that the experiment-system is in a defined state before and after an experiment conduction. Thereby, temporary data is removed and the standard configurations are reset.

- *Query and Discovery:* The cognitive process of knowledge discovery requires tools that help to query experiment-data, monitoring-data, and to put both into structural relationships. Furthermore, the computation of statistically sound performance metrics should be supported.

- *Adaptation:* Adaptation requires that the formal description of the experiment can be put in relationship to the obtained performance results and other contextual information of the experiment run.

Meeting these requirements for scientifically sound experimentation provides several benefits for the researcher. Most importantly, with a standardized experimentation description, other scientist can repeat the reported experiment, verify the results, and validate the conclusions. Thus, a certain degree of *repeatability* is provided and results from similar experiments across multiple testbeds can be compared more easily.

## 3 Large-Scale Wireless Testbeds

In this section we discuss a concrete implementation of the concepts from Section 2. For this, we briefly discuss in Section 3.1 related wireless testbeds (not exhaustive), afterwards in Section 3.2 we introduce the DES-Testbed. In Section 3.3 we present the user perspective of the testbed, and eventually in Section 3.4 the management system of the testbed, which is a realization of the scientific workflow of Section 2.

| | |
|---|---|
| CPU | 500 MHz AMD Geode LX800 |
| RAM | 256 MB DDR DRAM |
| WLAN interfaces | 2x MiniPCI IEEE802.11a/b/g AR5413 |
| | 1x USB IEEE802.11b/g Ralink RT2501 |
| Ethernet | 2 Ports (Via VT6105M) |

**Table 1:** *Specification of the mesh router of the DES-Node.*

## 3.1 Testbed initiatives

In the last decade, several initiatives deployed and run wireless multi-hop testbeds. In Europe, the ICT research framework declared testbed-based research as one of their priorities. The Future Internet Research and Experimentation (FIRE) initiative comprises testbed-based activities in this area [5]. The OpenLab project builds on existing wireless testbeds and develops a middleware to simplify cross-testbed experiments in wireless multi-hop networks [15]. Some of the projects use the *ORBIT management framework* (OMF) [8]. Specifically for wireless sensor networks, the WISEBED project [16] achieved a federation of wireless multi-hop sensor networks across Europe [17]. Additionally, several wireless mesh testbeds have been deployed in the scope of the EU OPNEX project [18]. The DES-Testbed was part of the WISEBED and OPNEX projects, and the largest involved wireless multi-hop testbed in the latter project.

## 3.2 The DES-Testbed

The *Distributed Embedded Systems Testbed* (DES-Testbed) is a hybrid wireless network located on the campus of the Freie Universität Berlin. Currently, 135 DES-Nodes are available. It is hybrid in a way, that a DES-Node (see Figure 4) consist of a wireless mesh router equipped with multiple IEEE 802.11a/b/g radios and an MSB-A2 sensor node [19]. Thus, a WMN based on IEEE 802.11 technology, called DES-Mesh, and a WSN, called DES-WSN, are operated in parallel. The specification of the wireless mesh router is given in Table 1, and for the MSB-A2 sensor node in Table 2.

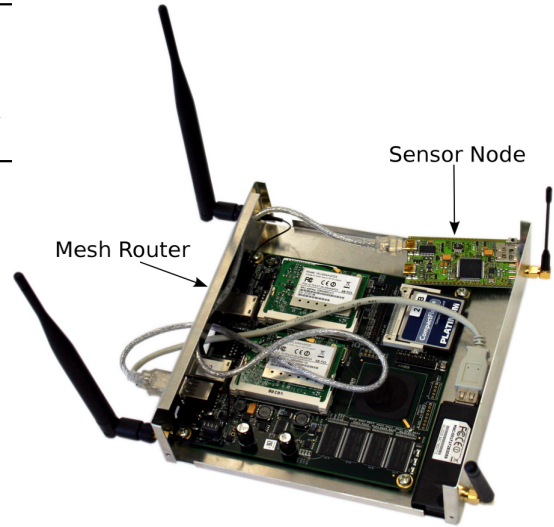

**Figure 4:** *A DES-Node comprising the multi-radio mesh router and the MSB-A2 sensor node. The mesh router is equipped with 3 wireless interfaces and the sensor node with one wireless interface.*

| | |
|---|---|
| Microcontroller | NXP Semiconductors LPC2387 |
| CPU Frequency | up to 72 MHz |
| RAM | 98 KiB |
| Flash | 512 KiB |
| Transceiver | Chipcon CC1100 at 868 MHz |

**Table 2:** *Specification of the MSB-A2 wireless sensor node.*

The DES-Nodes are scattered in an irregular topology across four buildings on the campus. Most of the nodes are deployed inside offices and some are outdoor at walls of the buildings. DES-Portal is the server of the testbed and provides central control capabilities. For this, the DES-Portal is connected to all DES-Nodes via an additional Ethernet backbone, which is only used for maintenance and management. As mobility is an important aspect for the research on WMNs and WSNs, Android-based smart phones mounted on Lego NXT robots serve as a mobile clients to the testbed [20].

## 3.3 User perspective

An important design goal of the DES-Testbed was to support the scientist throughout all steps of the scientific workflow for their experimental studies. The first crucial challenge was to provide convenient and reliable tools to access the testbed and the network nodes for experiments. The easiest way is to enable remote experimentation, which means that the scientist can access the testbed, i.e., he can create and schedule experiments remotely over the Internet. Remote experimentation is also an important requirement in order to attract external researchers and participation in international projects. The users of the DES-Testbed can access the testbed remotely via the DES-Portal.

A web front-end allows to design, formalize, and finally schedule experiments. A domain specific language for wireless network experiments supports the scientist in the process of formalizing the experiment. The language defines actions, that are executed at the network nodes during the experiment and their timings. The resulting structured experiment description has several advantages for the following steps of the scientific workflow. For the execution of the experiment, the final experiment description can be automatically executed on the testbed without further intervention of the scientist. Additionally, replications using the same experiment description can be easily scheduled thus improving the statistical reliability of the results. Based on the results of the feedback and analysis phase, the experiment can be easily adapted by changing the original experiment description.

### 3.4 Testbed Management System (TBMS)

The *DES-Testbed Management System* (DES-TBMS) comprises tools to support the design, automated execution, and analysis of experiments [7]. The software framework consists of six components, each dedicated to a specific task in the experiment workflow. The roles of the components and their relationships to the scientific workflow discussed in Section 2.2 are depicted in Figure 5.

DES-Cript is a *domain specific language* based on XML, which defines and describes network experiments in a holistic way [21]. As the structure of the underlying network is abstracted, DES-Cript is not limited to the DES-Testbed. Each DES-Cript file contains a general information section, followed by the available network nodes assigned into groups with particular roles. Next, actions are assigned to a group or individual nodes. As existing DES-Cript experiment descriptions can be edited and reused, they can be used to isolate critical parameters. Moreover, DES-Cript files provide a well-defined experiment documentation without any further effort. DES-Exp provides an experiment manager which is responsible for the scheduling and execution of experiments defined in the DES-Cript experiment description. DES-Exp also assures a defined state at the beginning of each experiment and its replications. DES-Web provides a web interface to DES-Exp, which allows to create, modify, and schedule experiments using DES-Cript. The network monitoring tool DES-Mon is based on SNMP and retrieves the network state from the DES-Nodes. DES-Mon collects data from the wireless interfaces, the kernel routing table, ETX link quality information [22], and data from the sensor nodes.

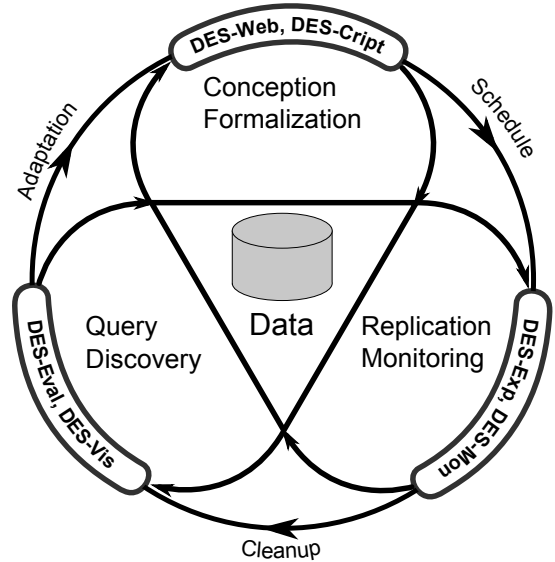

**Figure 5:** *The scientific workflow for empirical research as presented in Section 2.2 and the corresponding components of the DES-TBMS.*

DES-Vis is a 3D-visualization tool based on the JavaView framework [23] to display gathered data obtained from experiments or to show the current state of the network. For routing algorithms it can display the existing links between the network nodes or color the links and nodes for the evaluation of channel assignment algorithms. The evaluation tool DES-Eval enables the post-processing of the experiment results supporting workflows for an automatic evaluation process. It provides several configurable input, processing, and output modules. In addition, new modules can easily be written and integrated. Existing modules support input from log files or database records, statistical analysis using R [24], and output as plotted graphs.

## 4 Analysis of the Performed Experiments

In this section we present and discuss the performed experiments on the DES-Testbed in the years 2010 and 2011. Since a testbed is basically a *tool* for research, it is important to evaluate its usefulness for this purpose. For this, metrics are needed, however, it is difficult to elaborate appropriate ones. The most important metric in the evaluation of a scientific tool is the *scientific output* such as results of studies in the form of publications. Additionally, time dependent qualitative metrics such as the load of the testbed, the usability, and the efficiency are also helpful and, in comparison to the scientific output, simple to obtain.

### 4.1 Scientific output

The scientific output based on the DES-Testbed is summarized in Table 3 for the years 2010 and 2011. During the mentioned time interval six different projects used the testbed for experimentation. The projects were funded by BMBF[2], the European Union in FP 7, and the industry.

[2] Federal Ministry of Education and Research, Germany

More than 30 students worked with the testbed in this time interval and performed experiments required for their theses. The main scientific output is however, the academic publications in technical reports, international peer-reviewed conferences, workshops, and journals. In the years 2010 and 2011, a total of 25 papers have been published. The publications report experiment results, the framework of the DES-Testbed and its components, and the scientific method of the DES-Testbed.

| | |
|---|---:|
| Projects that used the testbed | 6 |
| Number of theses (B.Sc., M.Sc., etc.) | 32 |
| Total number of publications | 25 |
| - Conference papers | 15 |
| - Workshop papers | 5 |
| - Journal articles | 3 |
| - Technical reports | 2 |

**Table 3:** *Scientific output based on the DES-Testbed during the years 2010 and 2011.*

### 4.2 Experiments and experimenters

In this section we analyze the users of the testbed. The users of the DES-Testbed can be classified using two categories: i) status and ii) affiliation.

In the status category we have undergraduate and graduate *students* (B.Sc. and M.Sc.), and *PhD* candidates. This distinction is important, since the work style and the complexity of the research of PhDs and students differ. The PhDs work mostly independently, usually have a broader scientific background, and therefore have a better understanding of the scientific evaluation of experiments. Students may have none or minor experience with experimentation and scientific analysis (in a testbed environment). They do not work independently, and their experiments are typically part of a larger study directed by their supervisors.

For the affiliation, we distinguish between *internal* and *external* users. The internal users work geographically close to the testbed deployment area thus having physical access to the environ-ment of the testbed and its nodes. The external users are affiliated with other research groups that are partners of a mutual research project. External users are usually working far from the testbed (in other cities or countries) and usually do not have a physical access to the testbed. The external users therefore access the testbed and its nodes via the web-based experimentation interface (see Section 3.3).

This distinction should not be an important difference, since the typical workflow of running experiments is holistically supported by the remote experimentation and therefore the users do not require physical access to the testbed. However, it may be important to have a grasp of the testbed environment and to see the locations of the nodes. This can be helpful in interpreting the experiment results.

In Table 4 we summarized the information about the users for 2010 and 2011, respectively. The total number of users increased from 20 to 30 from 2010 to 2011. Most users of the testbed are internal students followed by internal PhDs. The smallest group in both years are external users. Although the number of external users is doubled in 2011, it represents only about 13% of the total users.

|  | 2010 | 2011 |
|---|---|---|
| Number of users | 20 | 30 |
| Internal users | 18 | 26 |
| - Students | 12 | 20 |
| - PhDs | 6 | 6 |
| External users | 2 | 4 |

**Table 4:** *The experimenters of the DES-Testbed distinguished between internal and external users.*

## 4.3 Performed experiments

Figure 6 depicts three different kind of information about the performed experiments on the DES-Testbed. i) the availability of the individual testbed nodes, ii) the distribution of the experiment runtimes, and iii) the number of testbed nodes used in experiments.

### 4.3.1 Availability of DES-Nodes

The availability of nodes, as depicted in Figure 6a and Figure 6b, shows to what degree the DES-Nodes were available for experimentation. In 2010 the average availability of each node is 98%, meaning that the nodes had a downtime of only 170.82 hours in average, i.e., 8 days in the whole year.

This performance is further improved in 2011 in which the average availability of nodes could be raised to 99.5%, which results in an average downtime of 2 days during the whole year. The main reasons for this improvement can be explained with two important developments. First, the testbed has been extended with additional nodes especially in the year 2010. In the first weeks after deployment, the office users are not used to the nodes yet and may accidentally unplug the power supply or the network cable, which leads to unexpected downtimes. The second reason is that over the time we continuously improved the software watchdog system running on the testbed nodes. This lead to automatic reboots, when the node was unable to connect to its NFS-mounted root filesystem or to the testbed server for a longer period of time.

### 4.3.2 Experiment runtimes

In Figure 6c and Figure 6d the distribution of the experiment runtime is depicted. In 2010 we have three interesting areas. The first observation is that a lot of experiments lasted only up to 1 hour. Second a healthy distribution of experiment runtimes between 2 hours and 11 hours exists. And finally, the remaining part of the graph, which embraces also the long-term experiments lasting longer than 24 hours.

The distribution of the experiment runtime differs in 2011. We do not have a mode of 1 hour experiments. Instead we have a healthy (normal) distribution of the experiment runtimes between 1 and 20 hours. And eventually the remaining part of the graph depicts the long-term experiments.

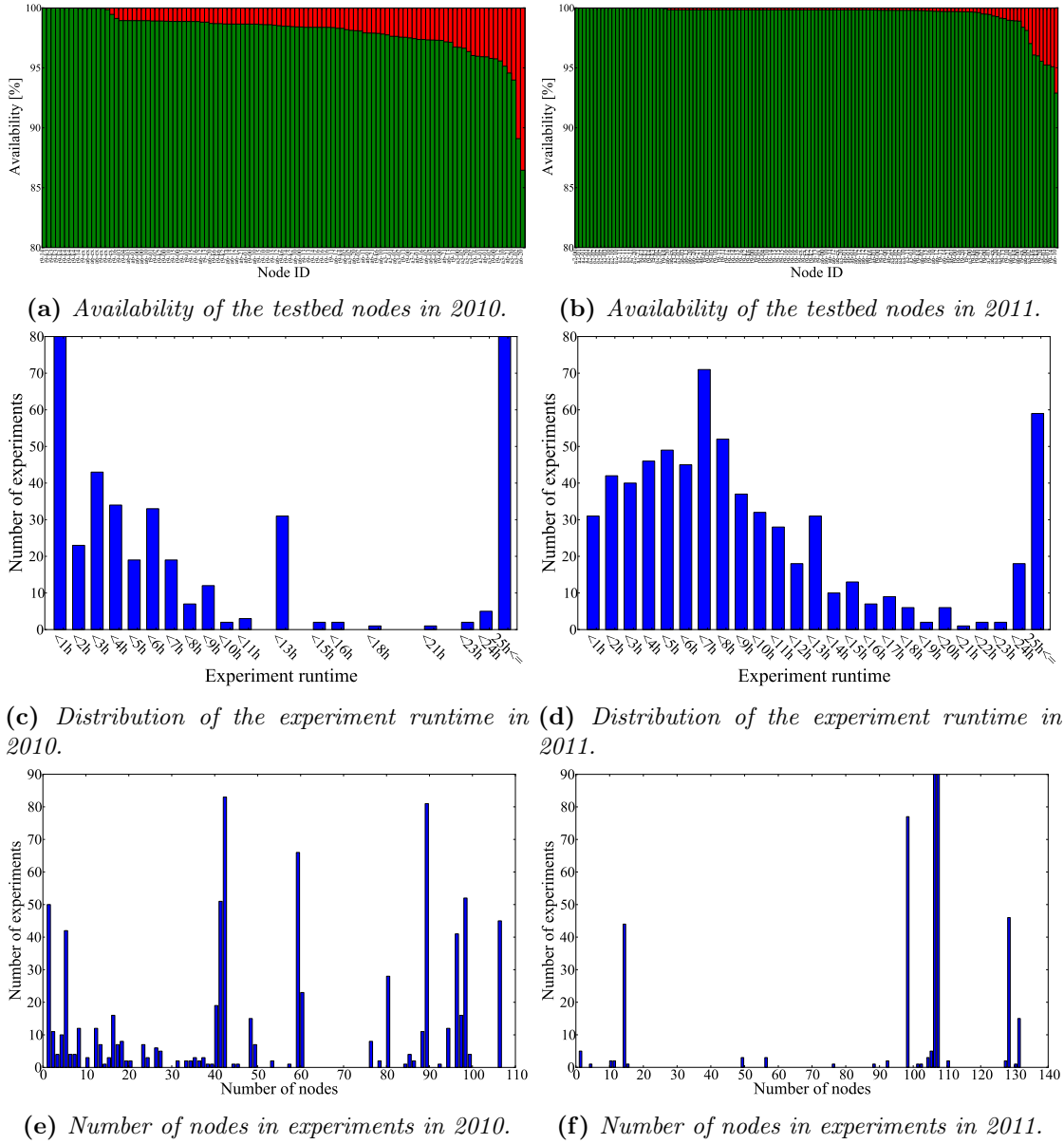

**Figure 6:** *The statistics about the performed experiments during 2010 and 2011. (a), (b) Show the availability of the individual nodes of the testbed. (c), (d) Show the distribution of the experiment runtimes. (e), (f) Show the involved number of nodes in the experiments.*

The main difference between the both years is that less short-term (less than one hour) experiments have been performed in the year 2011. This can be credited to our experience, with which we have learned a lot about the art of experimentation in the testbed in 2010 and could also teach this to our students. Also, many basic services, such as gathering the network topology, modules for node-to-node messaging, and interface settings are now in a stable state and are heavily reused by new experimenters. This process lead to less trial-and-error experiments and in general, in less time spent on debugging, but on actual useful experiments.

### 4.3.3 Number of nodes in experiments

In Figure 6e and Figure 6f the distribution of the number of nodes in the performed experiments is depicted. We witness a difference in the number of testbed nodes involved in experiments. In 2010 the number of nodes in experiments vary between 1 and 106 nodes and in 2011 between 1 and 131 nodes. The reason for the higher number in 2011 is due to the fact that we extended the testbed.

However, more interesting is the fact that the frequency of the number of nodes in the experiments differ strongly. In 2010 we observe a more uniform distribution of the number of nodes than in 2011. This may be due to the fact that experiments were run per building, i.e., all nodes in one building. In 2011 the number of nodes used in experiments look different. Experiments with 14, 98, 106, and 107 nodes occur most often, and the peak is at 106. The reason for this is that a lot of experiments were run with all available nodes.

Table 5 depicts a summary of the performed experiments on the DES-Testbed in 2010 and 2011, respectively. The number of performed experiments increased from 407 to 661 and at the same time the average duration of the experiment runtime. This proves for a high load of the testbed during both years, which is also underlined by the scientific output. At the same time this proves our testbed concept and architecture as very robust. Although, the number of users increased from 20 in 2010 to 30 in 2011, the average number of experiments per user decreased. This is a direct result of the higher number of users in 2011, since less time for experimentation for the individual user remained.

### 4.4 Type of experiments

Table 6 depicts an overview of the different kinds of experiments performed on the DES-Testbed. Additionally the total number of the experiment and the sum of the experiment runtime in hours is given. Since the testbed in not designed for a specific type of study or experiment, it can be used for all networking and communication topics. Although this sounds fairly obvious it is important to understand that different topics demand for various kind of support, i.e., for timing, distributing and gathering of information, and for the measurement.

### 4.5 Discussion

Based on our investigation, we can answer the original question of the value of a wireless multi-hop testbed in a very positive way. Since a testbed is a tool for scientific research our main metric for the evaluation is the scientific output. The number of publications shows indeed that experimental results are regarded as very important in the particular research community. The *production* of high-quality results on almost all layers of the network stack could only be achieved with the sophisticated management system. Researchers were forced to a sound experiment design and reaped the benefits of the provided tools such as automated experiment execution to enable sufficient experiment replications.

The statistics regarding the performed experiments and the testbed usage have shown, that a high demand for such experimental facilities exist. Throughout large-scale international research projects and university seminars, the testbed has seen hardly any idle time in the considered two years. External researchers using the testbed stress the importance to provide means of remote experimentation.

|                                       | 2010 | 2011 |
|---------------------------------------|------|------|
| Average uptime of nodes [%]           | 98   | 99.5 |
| Number of experiments                 | 407  | 661  |
| Max runtime of experiments [h]        | 132  | 875  |
| Average runtime of experiments [h]    | 8    | 11   |
| Max number of nodes in experiments    | 106  | 131  |
| Average number of nodes in experiments| 55   | 99   |
| Number of users                       | 20   | 30   |
| Average number of experiments per user| 39   | 22   |

**Table 5:** *Summary of the performed experiments in 2010 and 2011.*

|                     | 2010 |          | 2011 |          |
|---------------------|------|----------|------|----------|
| Topic               | Count| Time [h] | Count| Time [h] |
| Application Layer   | 10   | 20       | 70   | 721      |
| Channel Assignment  | 54   | 329      | 83   | 1116     |
| Localization        | 25   | 368      | 0    | 0        |
| MAC                 | 28   | 124      | 3    | 29       |
| Mobility            | 22   | 41       | 102  | 557      |
| Routing             | 75   | 451      | 362  | 4067     |
| Security            | 0    | 0        | 12   | 85       |
| Service Placement   | 6    | 72       | 0    | 0        |
| Tracking            | 12   | 58       | 0    | 0        |
| Transport Layer     | 0    | 0        | 18   | 216      |
| WSN*                | 66   | 276      | 78   | 598      |

**Table 6:** *The kind of experiments performed on the DES-Testbed. The count column depicts the number of experiments and the time column the total sum of experiment duration in hours. * Please notice that we consider all WSN expriments as one topic.*

## 5 Conclusions

We have shown that large-scale multi-hop testbeds are an important and useful tool for the research in wireless communications. However, the real work just begins after the initial deployment. A holistic understanding of the *experiment life cycle* is required to design a sophisticated management system that helps and guides the experimenter in all steps of the experiment by realizing a *scientific workflow*. The quantitative analysis of the DES-Testbed usage in the years 2010 and 2011 revealed that a large number of experiments have been performed and many results have been published. Thus, the high monetary and labor costs to deploy and maintain a large multi-hop testbed have been proven worth the investment.

## Acknowledgement

The work around the DES-Testbed at Freie Universität Berlin was an endeavor possible only by the effort of a lot of people, who contributed at different stages to the project. More than fifty students worked and used the testbed over the years. I thank all of them and would like to mention particularly by name: Dr. Achim Liers, Bastian Blywis, Felix Shzu-Juraschek, Oliver Hahm, Nicolai Schmittberger, Kaspar Schleiser, Pardeep Kumar, Qasim Mustaq.